 \documentclass[preprint2,longabstract]{aastex}

\usepackage{txfonts}
\usepackage{longtable}
\usepackage{rotating}
\usepackage{natbib}
\usepackage{graphicx}
\usepackage{graphics}
\usepackage{psfrag}
\usepackage{amssymb}
\bibpunct{(}{)}{;}{a}{}{,}
\def\Teff{$T_{\mathrm{eff}}$}

\def\vsini{\ensuremath{{\upsilon}\sin i}}
\def\kms{$\mathrm{km\,s}^{-1}$}
\def\ms{$\mathrm{m\,s}^{-1}$}
\def\kgs{$\mathrm{kg\,s}^{-1}$}

\def\Ro{\ensuremath{R_{\odot}}}

\def\Mo{\ensuremath{M_{\odot}}}

\def\ebv{\ensuremath{E(B-V)}}

\def\ergscm{erg\,s$^{-1}$\,cm$^{-2}$}
\def\mWm{mW\,m$^{-2}$}
\def\logR{\ensuremath{\log R^{\prime}_{\mathrm{HK}}}}
\def\loglx{\ensuremath{\log L_{\rm X}{\rm ([erg\,s^{-1}])}}}
\def\logleuv{\ensuremath{\log L_{\rm EUV}{\rm ([erg\,s^{-1}])}}}

\shorttitle{Suppressed Far-UV Stellar Activity and Low Planetary Mass Loss in the WASP-18 System}
\shortauthors{Fossati et al.}

\begin{document}


\title{Suppressed Far-UV Stellar Activity and Low Planetary Mass Loss in the WASP-18 System}
\altaffiltext{1}{Based on observations made with the NASA/ESA Hubble Space
Telescope, obtained from MAST at the Space Telescope Science Institute, which is
operated by the Association of Universities for Research in Astronomy, Inc.,
under NASA contract NAS 5-26555. These observations are associated with
program \#13859. Based on observations made with ESO Telescopes at the La Silla Paranal Observatory under programme ID 092.D-0587.}

\author{L. Fossati}
\affil{Space Research Institute, Austrian Academy of Sciences, Schmiedlstrasse 		6, A-8042 Graz, Austria}
\email{luca.fossati@oeaw.ac.at}
\and
\author{T. Koskinen}
\affil{Lunar and Planetary Laboratory, University of Arizona, 1629 East University Boulevard, Tucson, AZ 85721-0092, USA}
\email{tommi@lpl.arizona.edu}
\and
\author{K. France\altaffilmark{2}}
\affil{Laboratory for Atmospheric and Space Physics, University of Colorado, 600 UCB, Boulder, CO 80309, USA}
\email{kevin.france@colorado.edu}
\and
\author{P.~E. Cubillos}
\affil{Space Research Institute, Austrian Academy of Sciences, Schmiedlstrasse 		6, A-8042 Graz, Austria}
\email{patricio.cubillos@oeaw.ac.at}
\and
\author{C.~A. Haswell}
\affil{Department of Physical Sciences, The Open University, Walton Hall, Milton Keynes MK7 6AA, UK}
\email{C.A.Haswell@open.ac.uk}
\and
\author{A.~F. Lanza}
\affil{INAF -- Osservatorio Astrofisico di Catania, Via S. Sofia 78, I-95123 Catania, Italy}
\email{nuccio.lanza@oact.inaf.it}
\and
\author{I. Pillitteri\altaffilmark{3}}
\affil{INAF -- Osservatorio Astronomico di Palermo G.S. Vaiana, Piazza del Parlamento 1, I-90134, Palermo, Italy}
\email{pilli@astropa.inaf.it}

\altaffiltext{2}{Center for Astrophysics and Space Astronomy, University of\\ 			Colorado, 389 UCB, Boulder, CO 80309, USA.}
\altaffiltext{3}{Harvard-Smithsonian Center for Astrophysics, 60 Garden St.,\\ 			Cambridge, MA, USA.}

\begin{abstract}
WASP-18 hosts a massive, very close-in Jupiter-like planet. Despite its young age ($<$1\,Gyr), the star presents an anomalously low stellar activity level: the measured \logR\ activity parameter lies slightly below the basal level; there is no significant time-variability in the \logR\ value; there is no detection of the star in the X-rays. We present results of far-UV observations of WASP-18 obtained with COS on board of {\it Hubble Space Telescope} aimed at explaining this anomaly. From the star's spectral energy distribution, we infer the extinction (\ebv\,$\approx$\,0.01\,mag) and then the interstellar medium (ISM) column density for a number of ions, concluding that ISM absorption is not the origin of the anomaly. We measure the flux of the four stellar emission features detected in the COS spectrum (\ion{C}{2}, \ion{C}{3}, \ion{C}{4}, \ion{Si}{4}). Comparing the \ion{C}{2}/\ion{C}{4} flux ratio measured for WASP-18 with that derived from spectra of nearby stars with known age, we see that the far-UV spectrum of WASP-18 resembles that of old ($>$5\,Gyr), inactive stars, in stark contrast with its young age. We conclude that WASP-18 has an intrinsically low activity level, possibly caused by star--planet tidal interaction, as suggested by previous studies. Re-scaling the solar irradiance reference spectrum to match the flux of the \ion{Si}{4} line, yields an XUV integrated flux at the planet orbit of 10.2\,\ergscm. We employ the rescaled XUV solar fluxes to models of the planetary upper atmosphere, deriving an extremely low thermal mass-loss rate of 10$^{-20}$\,$M_{\rm J}$\,Gyr$^{-1}$. For such high-mass planets, thermal escape is not energy limited, but driven by Jeans escape.
\end{abstract}
%


\keywords{stars: individual (WASP-18) --- planets and satellites: individual (WASP-18b) --- stars: activity --- ultraviolet: stars}

\section{Introduction}\label{sec:introduction}
The discovery of close-in ($<$0.1\,au) giant planets (hot Jupiters) came as a surprise: their atmospheres were expected to escape rapidly under the effect of the strong stellar irradiation. Hubble Space Telescope ({\it HST}) ultraviolet (UV) observations have confirmed the presence of atmospheric escape at least on the hot Jupiters HD\,209458b, HD\,189733b, and WASP-12b as well as on the warm Neptune GJ436b \citep{vidal2003,fossati2010,linsky2010,haswell2012,bourrier2013,ehrenreich2015}. Despite these observations, thermal escape is not believed to significantly reduce the masses of typical hot Jupiters, such as HD\,209458b or HD\,189733b, within the lifetimes of their host stars \citep[e.g.][]{yelle2004}. The same may not hold true for extremely close-in giant planets with orbital periods of the order of one day, which are much more susceptible to mass loss \citep[e.g.][]{gaudi2017}. The impact of escape on the evolution of such planets, however, depends on their masses, which show a surprising diversity, ranging from Jovian-mass planets like WASP-12b to planets with several Jovian masses like WASP-18b. 

Indeed there is strong circumstantial evidence that mass loss may have a significant influence on the evolution of planets with orbital periods $P_{\rm orb}$\,$\sim$\,1\,day and masses $M_{\rm P}$\,$\lesssim$\,1\,$M_{\rm J}$. \citet{mazeh2016} present evidence for a profound lack of such planets in the population of known exoplanets. This cannot be explained by observational selection biases because close-in planets are easier to find than their longer period counterparts using both the RV and transit detection methods \citep[see e.g.][]{haswell2010}, and longer period analogues are much more plentiful. Furthermore, there are known planets at these short periods that have significantly lower masses, consistent with them being rocky planets. It is possible that some of these are the remnant cores of giants that have lost their outer layers through mass loss \citep{lecav2004}. 

The explanation that mass loss sculpts the demographics was suggested by \citet{szabo2011} who first identified the lack of short-period planets with $M_{\rm P}$\,$\lesssim$\,1\,$M_{\rm J}$ \citep[see also][]{davis2009}. An ingenious alternative explanation of this sub-Jovian desert in the demographics of known exoplanets was given by \citet{matsakos2016}. They show the demographics can be explained by the tidal circularization of those planets that acquire small pericenter eccentric orbits due to planet-planet scattering or Kozai--Lidov migration. Some such planets will be torn apart by differential gravitational forces because their pericenter is smaller than their Roche limit. Post-circularization, the surviving objects are neatly divided into giant planets above the sub-Jovian desert and rocky planets below it. Determining the mass-loss rates of planets above the upper boundary of the desert will allow us to ascertain whether mass loss does significantly contribute to the evolution of these planets and whether it explains or partially explains the existence of the sub-Jovian desert. The work we present here contributes to this effort and has ramifications for understanding the demographics of the Galaxy's population of planets.

\citet{hellier2009} reported the detection of a massive, short-period transiting hot Jupiter orbiting the mid-F-type \citep[F6V -- the stellar effective temperature \Teff\ is 6400$\pm$75\,K;][]{maxted2013} star WASP-18 (HD\,10069). The star is bright and close enough to have a measured {\it HIPPARCOS} \citep{perryman1997} parallax of 10.06$\pm$1.07\,mas, corresponding to a distance of 99.4$\pm$9.6\,pc \citep{vanLeeuwen2007}. The parallax, however, has been recently re-evaluated by combining the {\it TYCHO2} and {\it GAIA} data, leading to a value of 7.91$\pm$0.30\,mas, which corresponds to a distance of 126.4$\pm$4.6\,pc \citep{tgas}.

By combining the parameters obtained from their photometric and spectroscopic analyses with five different sets of stellar evolutionary tracks, \citet{southworth2012} estimated a stellar age of 0.4$^{+1.0}_{-0.4}$\,Gyr. \citet{maxted2013} performed a similar, though more in-depth, analysis setting an upper limit on the stellar age of 1.7\,Gyr. This value is consistent with the gyrochronological age of 1.1$^{+4.7}_{-0.6}$\,Gyr that they derived on the basis of the measurement of the stellar projected rotational velocity (\vsini) of 10.9$\pm$0.7\,\kms. We further discuss the age estimation at the light of the updated distance in Sect~\ref{sec:sed}.

The planet has a mass of 10.38$\pm$0.34\,$M_{\rm J}$, a radius of 1.163$\pm$0.055\,$R_{\rm J}$, and an orbital period of about 0.94\,days \citep{southworth2012}. Because of the close proximity \citep[about 0.02\,au;][]{southworth2012} to a rather hot star, the planet has a very high equilibrium temperature of 2413$\pm$44\,K \citep{southworth2012}. \citet{nymeyer2011} and \citet{sheppard2017} measured the planet's brightness temperature from {\it Spitzer} secondary eclipse observations, obtaining a value of about 3100\,K. From the analysis of {\it HST} secondary eclipse observations, \citet{sheppard2017} further concluded that WASP-18 hosts a metal rich atmosphere (about 200$\times$ solar).

Along with 25\% of known stars hosting massive, short-period planets \citep[e.g. WASP-12;][]{fossati2013,staab2017}, WASP-18 presents a rather low activity level of \logR\,$\approx$\,-5.15 \citep[][and references therein]{lanza2014}, which is just below the basal chromospheric flux level of main-sequence late-type stars of \logR\,=\,-5.1 \citep{wright2004}\footnote{The \logR\ value is a measure of the chromospheric emission in the core of the \ion{Ca}{2}\,H\&K resonance lines \citep{duncan1991} and allows intercomparison of the activity levels of FGK stars. See \citet{staab2017} for a graphical explanation.}. For comparison, the average solar \logR\ value is $-$4.902$\pm$0.063 \citep[95\% confidence level;][]{mamajek2008} and ranges between a minimum of about $-$5.0 and a maximum of about $-$4.8 along the activity cycle. During the Maunder minimum, the Sun was believed to have a \logR\ value of $-$5.102 \citep{baliunas1990,donahue1998,keil1998,radick1998,livingston2007}.

The high planetary mass and short orbital separation led several groups to study the WASP-18 system in search for star--planet interactions. \citet{miller2012} obtained 13 spectra of WASP-18 using the high-resolution spectrograph of the 6.5\,m Magellan Clay Telescope, looking for a correlation between the \logR\ parameter and the planetary orbital period. They also conducted simultaneous monitoring of the stellar X-ray flux with {\it Swift}. These observations showed no significant variability of the \logR\ parameter, in contrast to the expectation, and a low X-ray flux of \loglx\,$<$\,27.6, both indicative of low activity. 

To explain the anomalous activity of WASP-18, \citet{fossati2014} suggested the presence of a diffuse, translucent circumstellar gas cloud, such as that believed to surround WASP-12 \citep{haswell2012}. The latter hosts an inflated, short period hot Jupiter, which is subject to powerful mass-loss \citep{hebb2009,fossati2010,haswell2012,haswell2017}. \citet{haswell2012} and \citet{fossati2013} concluded that extrinsic absorption by material local to the WASP-12 system, and presumably escaping from the planet, is the most likely cause of the unexpected broad depression present at the cores of the \ion{Mg}{2}\,h\&k and \ion{Ca}{2}\,H\&K resonance lines \citep[note that this depression is always present, regardless of the planet's orbital phase;][]{haswell2012,nichols2015}. General support to the idea that circumstellar material may be present around stars hosting close-in planets has been given by \citet{cohen2011}, \citet{france2013}, \citet{lanza2014}, \citet{matsakos2015}, \citet{fossati2015c}, \citet{carroll2016}, and \citet{staab2017}. 

\citet{pillitteri2014} reported the results of {\it Chandra} X-ray observations of WASP-18, which led to a non-detection with an upper limit of \loglx\,$<$\,26.6. On the basis of the {\it GAIA} distance, the updated upper limit on the X-ray luminosity is \loglx\,$<$\,26.9. This finding is in stark contrast with the rather young age of the star. \citet{pillitteri2014} concluded that extrinsic absorption is most likely not present, because the absorption would attenuate only the soft X-ray spectrum, while the hard component should still be detectable, in contrast with the observations. \citet{pillitteri2014} argued therefore that the anomalously low activity of WASP-18 could be caused by the presence of the close-in, massive planet, which disrupts the dynamo generated in the thin stellar convective layer.

It is, in principle, possible to distinguish between these two scenarios (i.e., intrinsic low activity or extrinsic absorption) by observing and measuring the strength of low- and high-ionization chromospheric lines (e.g. \ion{C}{1}, \ion{C}{2}, and \ion{Si}{4}), comparing them with those of nearby stars that do not host close-in planets. This is because the strength of low-ionization features is controlled by the activity of the star and by the extrinsic absorption, while the strength of the high-ionisation lines is driven exclusively by stellar activity. To this end, we obtained a far-UV (FUV) spectrum of WASP-18 using the Cosmic Origins Spectrograph \citep[COS;][]{green2012} on the {\it HST} to detect and measure lines arising from the chromosphere and transition region. We present here the analysis of this COS observation and an in-depth modeling of the planetary upper atmosphere.

The ramifications of our work go beyond the understanding of the WASP-18 system itself. Stellar activity is directly related to the EUV flux \citep[e.g.][]{piters1997}, which plays a fundamental role in planetary atmospheric escape \citep[e.g.][]{lammer2003,koskinen2013a,koskinen2013b}, hence shaping planetary evolution \citep[e.g.][]{mordasini2012,beauge2013,owen2013,owen2017,haswell2017,jin2017}. Because many planet-hosting stars have a low activity level, it is crucial to carefully study the activity of fairly inactive stars to accurately infer their EUV fluxes. Nevertheless, given their intrinsically low levels of chromospheric emission, all but the nearest inactive late-type stars have proved difficult to study at UV wavelengths. The installation of COS renewed interest in observing low-activity stars in the UV \citep[e.g.][]{france2013,linsky2013,stelzer2013,fossati2015b}.

We describe in Sect.~\ref{sec:observations} the {\it HST} observations and the applied data reduction. We present results in Sect.~\ref{sec:results}, which we then discuss in Sect.~\ref{sec:discussion}. We present the conclusions in Sect.~\ref{sec:conclusions}.
\section{Observations and data reduction}\label{sec:observations}
We observed WASP-18 using the FUV channel of {\it HST}/COS. The observations were conducted on 2015 July 15 along two consecutive {\it HST} orbits, which covered the 0.09--0.36 planetary orbital phases \citep[ephemeris from][]{triaud2010}: entirely out of transit. The total exposure time was 4988\,seconds. We adopted the instrument settings of the COS WASP-13 observations described by \citet{fossati2015b}: FUV G140L grating at the 1105\,\AA\ setting, which provides a continuous spectrum over the 1128--2236\,\AA\ wavelength range at a spectral resolution of $R\sim$3000, in TIME-TAG mode.

\begin{figure*}[h!]
\includegraphics[width=\hsize,clip]{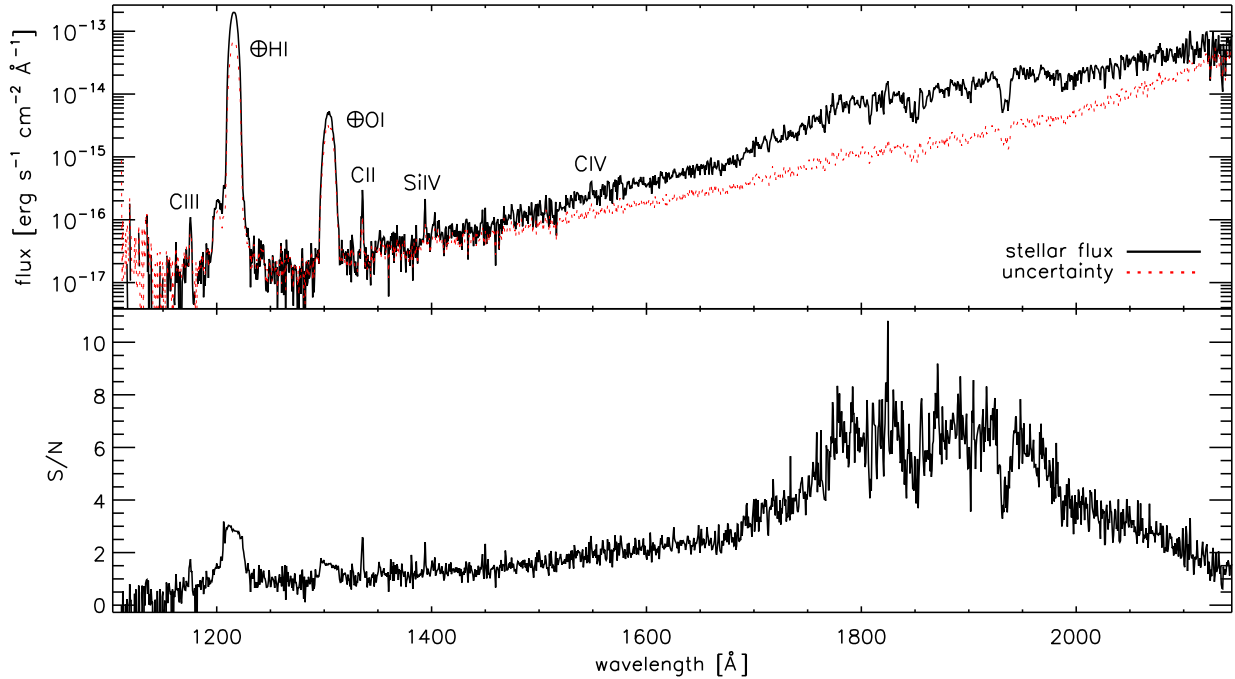}
\caption{Top panel: flux-calibrated co-added COS spectrum of WASP-18 (black solid line) and corresponding uncertainties (red dotted line). The strong emission lines at short wavelengths are caused by \ion{H}{1} and \ion{O}{1} geocoronal emission. The four detected and measured stellar features (see Sect.~\ref{sec:line_flux}) are marked. Bottom panel: signal-to-noise ratio (S/N) per resolution element of the co-added COS spectrum as a function of wavelength.}
\label{fig:whole-spectrum}
\end{figure*}

We retrieved the data, reduced with CALCOS V.3.0, from the MAST\footnote{\tt http://archive.stsci.edu/} archive and co-added the two flux-calibrated spectra, using the routine described by \citet{danforth2010}. The co-added FUV spectrum of WASP-18 is shown in Fig.~\ref{fig:whole-spectrum}. The two strongest emission lines are caused by geocoronal emission from atomic hydrogen and oxygen, while the rise in flux with increasing wavelength is due to the stellar photosphere.
%
\section{Results}\label{sec:results}
\subsection{Interstellar extinction}\label{sec:sed}
Ions abundant in the interstellar medium (ISM) absorb strongly at the wavelengths of their resonance lines, including some that are measurable in the COS spectrum. Consequently, we examine the ISM absorption along the line of sight of WASP-18.

We started by fitting synthetic fluxes, calculated with {\sc marcs} models \citep{marcs} and rescaled for the measured distance of 126.4$\pm$4.6\,pc \citep{tgas}, to the observed Johnson \citep{kharchenko2009}, TYCHO \citep{hog2000}, 2MASS \citep{cutri2003}, and {\it WISE} \citep{cutri2012} photometry, converted to physical units. We converted the photometry adopting the calibrations provided respectively by \citet{bessel1998}, \citet{grossmann1995}, \citet{vander}, and \citet{wright2010}. From the fit, we derived a stellar radius of 1.30$\pm$0.05\,\Ro\ and a low color excess \ebv\ of 0.023$\pm$0.009\,mag, where the error bars include the uncertainties in the stellar distance and photometry. The stellar radius is in agreement with the value of 1.255$\pm$0.028\,\Ro\ given by \citet{maxted2013}.

The slightly larger stellar radius, which is most likely caused by the adoption of a larger distance, has an impact on both planetary radius and brightness temperature. We therefore re-derived both of them, obtaining a significant increase in the planetary radius of 0.1\,$R_{\rm J}$ and a corresponding decrease in the brightness temperature of 130\,K, compared to the values given respectively by \citet{southworth2012} and \citet{nymeyer2011}.

An increase in the stellar radius leads to an increase in the inferred stellar luminosity, which may therefore have an effect on the estimated age of the star. We made use of the Bayesian PARAM\footnote{\tt http://stev.oapd.inaf.it/cgi-bin/param} tool \citep{dasilva2006}, employing the stellar evolutionary tracks by \citet{bressan2012} and as priors the stellar initial mass function by \citet{chabrier2001} and a constant star formation rate. Taking into account the {\it HIPPARCOS} distance, PARAM returned a stellar mass of 1.23$\pm$0.04\,\Mo, radius of 1.20$\pm$0.06\,\Ro, and age of 0.61$\pm$0.58\,Gyr, while by employing the {\it GAIA} distance, we obtained a mass of 1.27$\pm$0.04\,\Mo, radius of 1.26$\pm$0.05\,\Ro, and age of 0.84$\pm$0.76\,Gyr. The stellar radius derived on the basis of the {\it GAIA} distance is in agreement with those given by \citet{maxted2013} and derived here from the fitting of the spectral energy distribution. More importantly, these results indicate that the larger distance has no significant effect on the previously inferred stellar age estimate, further confirming that WASP-18 is a rather young star, in particular when considering that the main-sequence lifetime of a 1.2\,\Mo\ star is of the order of 4\,Gyr \citep{bressan2012}. 

We further derived the \ebv\ value from the Galactic extinction maps of \citet{amores2005} obtaining \ebv\,=\,0.008\,mag and from the Infrared Science Archive (IRSA)\footnote{\tt http://irsa.ipac.caltech.edu/applications/DUST/} dust maps obtaining a value of \ebv\,=\,0.0115$\pm$0.0012. This last value lies in between that derived from the fitting of the spectral energy distribution and from the maps of \citet{amores2005}. Although the star lies more than 100\,pc away from us, the low ISM absorption in the direction of WASP-18 is confirmed by the results of the {\it HIPPARCOS} mission, which indicate that along the WASP-18 line of sight, particularly within the first 50\,pc, the ISM is sparse \citep[see e.g. Fig.~1 of][]{frisch2011}.

We therefore used the maximum and minimum \ebv\ values (i.e. 0.023$\pm$0.009 and 0.008\,mag) to estimate the range of ISM column densities for \ion{H}{1}, \ion{C}{1}, \ion{C}{2}, \ion{C}{3}, \ion{C}{4}, \ion{Si}{4}, \ion{Mg}{2}, \ion{Ca}{2}, and \ion{Na}{1}. Following the results and considerations of \citet{fossati2015b,fossati2017}, we derived the column densities employing the $N_{\rm HI}$--\ebv\ conversion\footnote{Given the small \ebv\ value, we assumed that hydrogen is not in molecular form \citep[e.g.][]{rachford2002}.} provided by \citet{savage1979} and the ISM abundances and ionization mix given by \citet[][their model 2 -- Table~5]{frisch2003}. The results are listed in Table~\ref{tab:ebv-columnDensities}. We remark that the use of other $N_{\rm HI}$--\ebv\ conversions, such as those of \citet{diplas1994} or \citet{guver2009}, does not significantly affect the results \citep[e.g.][]{fossati2015b,fossati2017}.
\begin{table}[ht]
\caption[ ]{ISM column densities (in cm$^{-2}$). The column densities are given for the derived maximum and minimum \ebv\ values (see text). The uncertainties account for the error bar on \ebv, only.}
\label{tab:ebv-columnDensities}
\begin{center}
\begin{tabular}{l|cc}
\hline
\hline
\ebv    & 0.023$\pm$0.009 & 0.008 \\
Element & \multicolumn{2}{|c}{$\log N_{\rm X}$} \\
\hline
\ion{H}{1}  & 20.13$^{+0.14}_{-0.22}$ & 19.67 \\
\ion{C}{1}  & 13.21$^{+0.14}_{-0.22}$ & 12.75 \\
\ion{C}{2}  & 16.53$^{+0.14}_{-0.22}$ & 16.07 \\
\ion{C}{3}  & 15.05$^{+0.14}_{-0.22}$ & 14.59 \\
\ion{C}{4}  &  0.00$^{+0.00}_{-0.00}$ &  0.00 \\
\ion{Si}{4} & 10.41$^{+0.14}_{-0.22}$ &  9.95 \\
\ion{Mg}{2} & 14.71$^{+0.14}_{-0.22}$ & 14.25 \\
\ion{Ca}{2} & 12.77$^{+0.14}_{-0.22}$ & 12.53 \\
\ion{Na}{1} & 11.59$^{+0.14}_{-0.22}$ & 11.13 \\
\hline
\end{tabular}
\end{center}
\end{table}


Unfortunately, none of the stars analyzed by \citet{redfield2002,redfield2004} are close to the WASP-18 line of sight. However, there are a few stars that lie 20$^{\circ}$--50$^{\circ}$ away and for which \ion{Ca}{2} and \ion{Mg}{2} column densities have been measured. For these stars, \citet{redfield2002,redfield2004} listed maximum column density values of 11.0 and 14.5 for \ion{Ca}{2} and \ion{Mg}{2}, respectively, values that are consistent with those in Table~\ref{tab:ebv-columnDensities}, as these stars lie much closer to us than WASP-18. We also looked for hot, massive stars, lying close to us and to the WASP-18 line of sight. Hot stars facilitate direct measurements of the ISM column density for various ions. Within an angular distance of 10\,$^\circ$, there are two nearby hot, massive stars, $\kappa$\,Eri and $\phi$\,Eri, which lie, respectively, 155.8$\pm$3.6 and 47.1$\pm$0.3\,pc from us \citep{vanLeeuwen2007}. Because of the distance ($\sim$156\,pc) similar to that of WASP-18 ($\sim$126\,pc), $\kappa$\,Eri is the most interesting of the two stars. For this star, \citet{welsh1994} and \citet{welsh1997} reported ISM column densities for \ion{H}{1}, \ion{Ca}{2}, and \ion{Na}{1} of $\log(N_{\rm X}$[cm$^{-2}$])\,=\,19.58, 12.29$\pm$0.05, and 11.27, respectively. They also listed a \ebv\ value of 0.02\,mag. These ISM column density values agree well with those given in Table~\ref{tab:ebv-columnDensities} and in particular with those based on the lower \ebv\ value of 0.008\,mag.

For the elements considered here, \ion{C}{2}, \ion{Si}{2}, \ion{Mg}{2}, \ion{Ca}{2}, and \ion{Na}{1} are the main ionization stages in the local ISM. At the low \ebv\ value of WASP-18, almost all of the carbon in the diffuse ISM is in the singly ionized state \citep{snow2006}. \ion{C}{4} is not typically observed in the local ISM, which is why Table~\ref{tab:ebv-columnDensities} lists a null column density \citep{frisch2003,frisch2011}, though a single detection of \ion{C}{4} in the local ISM has been reported by \citet{welsh2010}.

From the values listed in Table~\ref{tab:ebv-columnDensities}, we conclude that only \ion{H}{1} and \ion{C}{2} are significantly affected by ISM absorption (i.e. $\log N_{\rm X}$\,$>>$\,15) and that \ion{Mg}{2} and \ion{C}{3} may be affected too, though high-resolution UV spectra would be necessary to identify the ISM absorption features. We also remark that the core of the \ion{Ca}{2}\,H\&K and \ion{Na}{1}\,D lines of WASP-18, observed with the high-resolution UVES spectrograph of the ESO Very Large Telescope (VLT), does not present any signature of ISM absorption \citep{fossati2014}. We also used this optical spectrum to look for features of diffuse interstellar bands, finding none. We can therefore safely conclude that the discrepancy between the young stellar age and the low activity is not caused by extrinsic ISM absorption, in agreement with the conclusions of \citet{fossati2014} and \citet{pillitteri2014}.
\subsection{Intrinsic line emission}\label{sec:line_flux}
\begin{figure*}
\includegraphics[scale=1.]{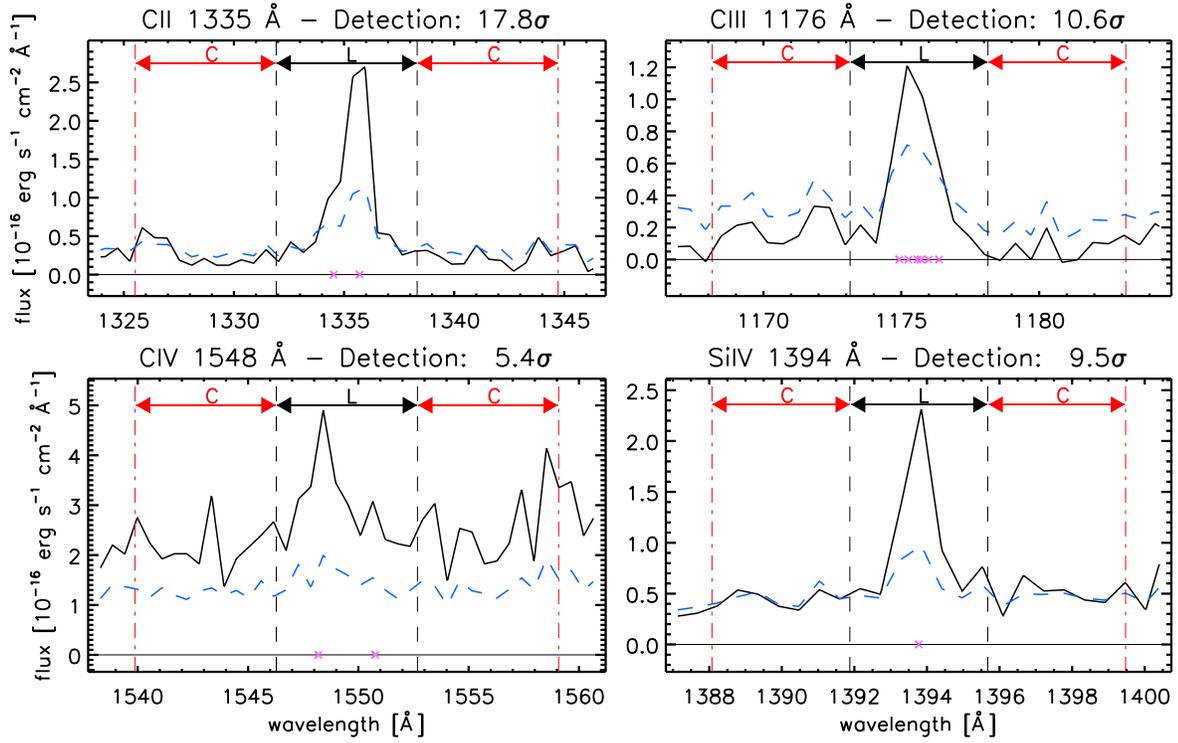}
\caption{COS spectrum of WASP-18 in the region of the four analyzed features. The black solid line and the blue dashed line show, respectively, the stellar flux and its uncertainty in units of 10$^{-16}$\,erg\,s$^{-1}$\,cm$^{-2}$\,\AA$^{-1}$. The crosses at the bottom of each panel indicate the expected position of the components of the stellar emission lines. The vertical black dashed lines mark the regions used to measure the integrated line fluxes, further indicated by black arrows and an L on the top of each panel. The red dotted--dashed lines mark the two regions for each feature used to derive the integrated continuum flux, further indicated by red arrows and a C on the top of each panel. The detection significance in units of the standard deviation of the background is given above each panel.}
\label{fig:lines_w18}
\end{figure*}

We inspected the COS spectrum and found just four clearly detected stellar emission features, namely the \ion{C}{3} multiplet at $\sim$1176\,\AA, the \ion{C}{2} doublet at $\sim$1335\,\AA, the \ion{Si}{4} line at $\sim$1394\,\AA, and the \ion{C}{4} doublet at $\sim$1548\,\AA\  (see Fig.~\ref{fig:whole-spectrum}). To measure the fluxes of these features, we employed the procedure adopted for WASP-13 \citep{fossati2015b}. The wavelength regions considered for each feature are shown in Fig.~\ref{fig:lines_w18}, while the resulting integrated fluxes and line detections are listed in Table~\ref{tab:wasp18-linefluxes}.
\begin{table*}[ht]
\caption[ ]{Results obtained from the analysis of the considered FUV emission lines detected in the COS spectrum of WASP-18. Column (1) gives the ion and approximate wavelength. Column (2) gives the wavelength range adopted to derive the integrated line flux. Columns (3)--(5) list the integrated line, continuum, and continuum-subtracted integrated fluxes for each feature in units of 10$^{-16}$\,\ergscm. The last column lists the $\sigma$ detection.}
\label{tab:wasp18-linefluxes}
\begin{center}
\begin{tabular}{l|c|ccc|c}
\hline
\hline
Feature & Line Wavelength & Line & Cont. & Cont. Sub. & $\sigma$ Detection \\
        & Range           & Flux & Flux  & Line Flux  &                    \\
        & [\AA]           & \multicolumn{3}{c|}{[10$^{-16}$ \ergscm]}    & \\
\hline
\ion{C}{2}  -- 1335\,\AA & 1331.92--1338.30 &  5.76$\pm$0.21 &  1.58$\pm$0.11 &  4.18$\pm$0.24 & 17.8 \\
\ion{C}{3}  -- 1176\,\AA & 1173.17--1178.12 &  2.37$\pm$0.14 &  0.63$\pm$0.09 &  1.74$\pm$0.17 & 10.6 \\
\ion{C}{4}  -- 1548\,\AA & 1546.29--1552.67 & 18.62$\pm$0.51 & 14.91$\pm$0.45 &  3.71$\pm$0.68 &  5.4 \\
\ion{Si}{4} -- 1394\,\AA & 1391.91--1395.66 &  3.76$\pm$0.17 &  1.79$\pm$0.12 &  1.98$\pm$0.21 &  9.5 \\
\hline
\end{tabular}
\end{center}
\end{table*}


Figure~\ref{fig:lines_w18} and Table~\ref{tab:wasp18-linefluxes} indicate that all four lines are clearly detected. Note that the background level around the \ion{C}{4} doublet at $\sim$1548\,\AA\ is much higher than that of the other three lines because the stellar photospheric flux rises significantly between 1400 and 1500\,\AA. From the COS spectrum, we also derived an upper limit on the strength of the \ion{C}{1} multiplet at $\sim$1657\,\AA, obtaining a value of $<$\,5.7$\times$10$^{-16}$\,\ergscm.

\begin{figure}
\includegraphics[width=\hsize]{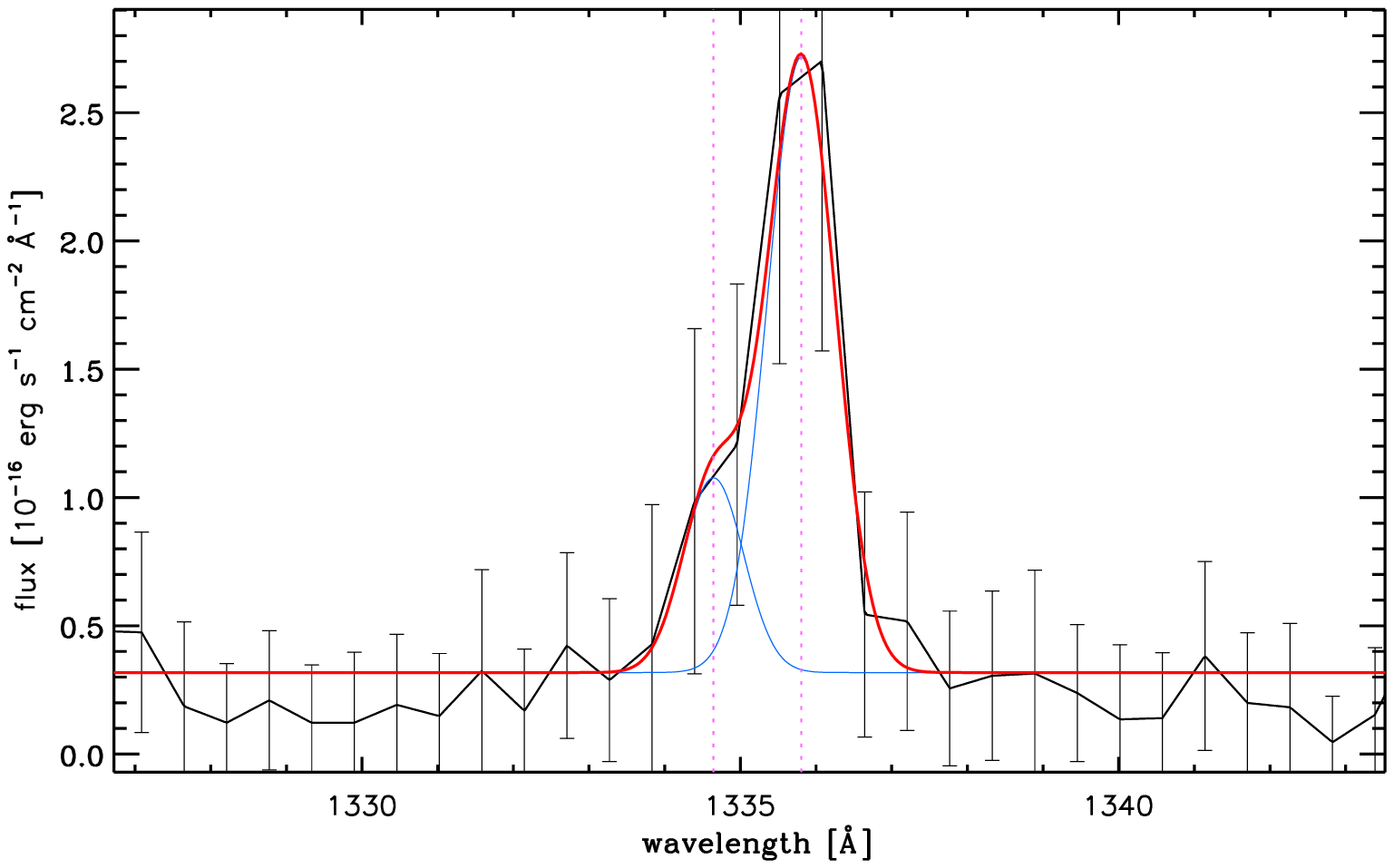}
\caption{Fit of a double-peak Gaussian to the observed profile of the \ion{C}{2} doublet at $\sim$1335\,\AA. The black line with error bars shows the observed COS spectrum, the blue lines present the two Gaussian components, while the thick red line is the sum of the two Gaussian functions. The vertical dotted lines show the position of the \ion{C}{2} 1334 and 1335\,\AA\ features in the laboratory rest frame.}
\label{fig:c2fit}
\end{figure}

On the basis of the results presented in Sect.~\ref{sec:sed}, we can safely conclude that the fluxes measured for \ion{C}{4} and \ion{Si}{4} are unaffected by ISM absorption. The same applies to the measured \ion{C}{3} multiplet because this does not arise from ground-state transitions (i.e. no resonant lines). On the other hand, the flux measured for \ion{C}{2} is almost certainly reduced by ISM absorption. However, thanks to the high spectral resolution and good data quality, we can separate the \ion{C}{2} doublet at $\sim$1335\,\AA\ into the two components (i.e. 1334 and 1335\,\AA), measuring them separately. We did this by fitting the observed profile with a double-peak Gaussian, taking the background into account (Fig.~\ref{fig:c2fit}). This is useful because \ion{C}{2} ISM absorption affects mostly the 1334\,\AA\ feature, which rises from a resonance transition, while the 1335\,\AA\ feature is only marginally affected as a result of the large ISM \ion{C}{2} column density. For the 1335\,\AA\ line, we obtained a continuum-subtracted integrated flux of 3.83($\pm$0.16)$\times$10$^{-16}$\,\ergscm, which, because of the possible contamination of ISM absorption, should be considered as a lower limit. The presence of \ion{C}{2} emission is already an indication that extrinsic absorption, in particular from circumstellar material, is probably not at the origin of the anomalously low activity level.
%
\section{Discussion}\label{sec:discussion}
\subsection{Comparison with WASP-13}\label{sec:compw13}
In the COS spectrum of WASP-13, we obtained tentative 2--4$\sigma$ detections of the \ion{Si}{4} line at 1403\,\AA\ and the \ion{C}{1} multiplets at 1560 and 1657\,\AA\ that are not detected in the FUV spectrum of WASP-18. The \ion{Si}{4} line at 1403\,\AA\ is intrinsically weaker than the \ion{Si}{4} 1394\,\AA\ feature. We attempted to detect this weaker feature in WASP-18, but obtained just a $\approx$3$\sigma$ detection, which a visual inspection of the spectrum suggests being possibly spurious. The non-detection of \ion{C}{1} could be attributed to the fact that the \ion{C}{1} lines may not be strong enough to emerge over the stellar photospheric flux and/or by the fact that the \ion{C}{1} lines are intrinsically weaker for WASP-18 than for WASP-13, hence lost in the noise. This last possibility would in turn imply either the presence of a hotter or thicker (i.e. more emitting power or volume) chromosphere compared to that of WASP-13. For a meaningful comparison of the FUV spectra of WASP-13 and WASP-18, it is important to note that the spectrum of WASP-13 is of much lower quality than that of WASP-18, as indicated, for example, by the fact that for WASP-13, we could not separate the two components of the \ion{C}{2} doublet at $\sim$1335\,\AA.

To identify which of these two possibilities is correct, we compare the flux ratios obtained for the lines detected and measured for both stars. Because of ISM absorption, we cannot rely on the fluxes measured for \ion{C}{2}\footnote{Note that the \ion{C}{3} multiplet is not detected in the spectrum of WASP-13.}, which is why we compare the flux ratios between the \ion{C}{4} and \ion{Si}{4} features, which form in the transition region. We note that the formation temperature of \ion{C}{4} is $\sim$107\,kK, while that of \ion{Si}{4} is $\sim$62\,kK \citep[see e.g. Table~6 of][]{pagano2004}. For WASP-18, we obtained a \ion{C}{4}/\ion{Si}{4} flux ratio of 1.87$\pm$0.35, while for WASP-13, the values reported by \citet{fossati2015b} lead to a flux ratio of 3.37$\pm$0.63. These line flux ratios give the ratios of the emission measures at the respective line formation temperatures, assuming thermodynamic equilibrium. This implies that the emission measure at higher temperature is relatively lower in WASP-18 than in WASP-13, thus indicating a lower non-radiative heating of the upper atmosphere of the former. This further suggests that the transition region of WASP-13 is probably hotter or thicker than that of WASP-18, allowing us to conclude that the \ion{C}{1} multiplets at 1560 and 1657\,\AA\ are probably intrinsically weak and therefore not able to emerge significantly above the photospheric flux. This is interesting because in \citet{fossati2015b} we concluded that WASP-13 has a chromosphere and transition region similar to those of other inactive solar-type stars, very similar to those of the Sun and $\alpha$\,Cen\,A. This indicates that WASP-18 may indeed have an intrinsically low activity level, consistent with the lack of detection in the X-rays \citep{pillitteri2014}. To confirm this, we compare then the FUV line flux ratios with those of other stars of similar temperature.
\subsection{WASP-18's chromospheric activity}\label{sec:activity}
We inspected the {\it HST} Spectroscopic Legacy Archive (HSLA)\footnote{{\tt http://archive.stsci.edu/hst/spectral\_legacy/}} and the StarCAT catalog \citep{ayres2010} looking for FUV spectra of main-sequence stars of various ages and activity levels, and with a spectral type ranging between G0 and F3 (WASP-18's spectral type is F6V) to measure their \ion{C}{3}/\ion{C}{4} and \ion{C}{2}\footnote{In this section, \ion{C}{2} stands for the \ion{C}{2} line at 1335\,\AA.}/\ion{C}{4} flux ratios to be then compared to those of WASP-18. The comparison stars are listed in Table~\ref{tab:comp.stars}, where we also give their age and \logR\ values. All stars but HD\,220657 were already considered as comparison stars for WASP-13 by \citet{fossati2015b}, and we adopt the same age and \logR\ values employed there.
\begin{table}[ht]
\caption[ ]{Comparison stars. The spectral types (column (2)) are taken from Simbad. Columns (3), (4), and (5) list the stars' distance, \logR\ and age (in Gyr). The stellar distances are taken from TGAS \citep{tgas}, while, unless otherwise stated, the stars' \logR\ values and ages are taken from \citet{barnes2007}. If multiple \logR\ were available, we opted for the lowest one. The typical uncertainty on the \logR\ values is of 0.05--0.1\,dex, while for the age the typical uncertainty is of 500\,Myr \citep[e.g.][]{casagrande2011}.} 
\label{tab:comp.stars}
\begin{center}
\begin{tabular}{l|c|c|c|c}
\hline
\hline
Star & Spectral & Distance & \logR\ & Age   \\
     & Type     & (pc)     &        & (Gyr) \\
\hline
HD\,39587   & G0V  &   8.7$\pm$0.1  & $-$4.43     & 0.29      \\ 
HD\,97334   & G0V  &  21.9$\pm$0.2  & $-$4.42     & 0.55      \\ %
HD\,209458  & G0V  &  49.6$\pm$1.9  & $-$4.97$^e$ & 4.00$^b$  \\
HD\,142373  & G0V  &  15.9$\pm$0.1  & $-$5.11$^a$ & 7.40$^d$  \\
HD\,33262   & F9V  &  11.6$\pm$0.1  & $-$4.65$^c$ & 0.25$^d$  \\
HD\,220657  & F8IV &  52.2$\pm$0.5  & $-$4.54$^c$ & 0.85$^d$  \\
\hline
\end{tabular}
\end{center}
$^a$ -- From \citet{mamajek2008}.\\
$^b$ -- From \citet{melo2006}.\\
$^c$ -- From \citet{pace2013}.\\
$^d$ -- From \citet{casagrande2011}.\\
$^e$ -- From \citet{figueira2014}.\\
\end{table}


Unfortunately, all these comparison stars are cooler than WASP-18, with F8 being the earliest spectral type. This is not an ideal situation, in particular because the surface convective layers of stars corresponding to mid-F spectral types become progressively thinner, hence affecting the chromospheric and coronal structures. We searched the IUE archive for FUV high-resolution spectra of stars with spectral type earlier than F8, but we were able to detect measurable features just for HD\,11443, which is an evolved star (F5III) and hence unsuitable. The only way to fully solve this problem would be to obtain {\it HST} FUV spectra of nearby main-sequence stars of known age in the F8--F3 spectral range. Such observations would be crucial to improve our understanding of stellar structure and activity, and not just to provide appropriate comparisons for WASP-18.

The spectra of HD\,142373, HD\,220657, HD\,33262, HD\,39587, and HD\,97334 were obtained with STIS (E140M grating), and we downloaded them from the StarCAT catalog \citep{ayres2010}, while that of HD\,209458 was obtained with COS and downloaded from HSLA \citep{france2010}. The COS spectra obtained with the G130M and G160M gratings have been co-added with the same tool employed to co-add the WASP-18 COS spectra. Before measuring the line fluxes, all spectra were convolved with a Gaussian to match the spectral resolution of the COS spectrum of WASP-18.

\begin{figure*}
\includegraphics[scale=1.]{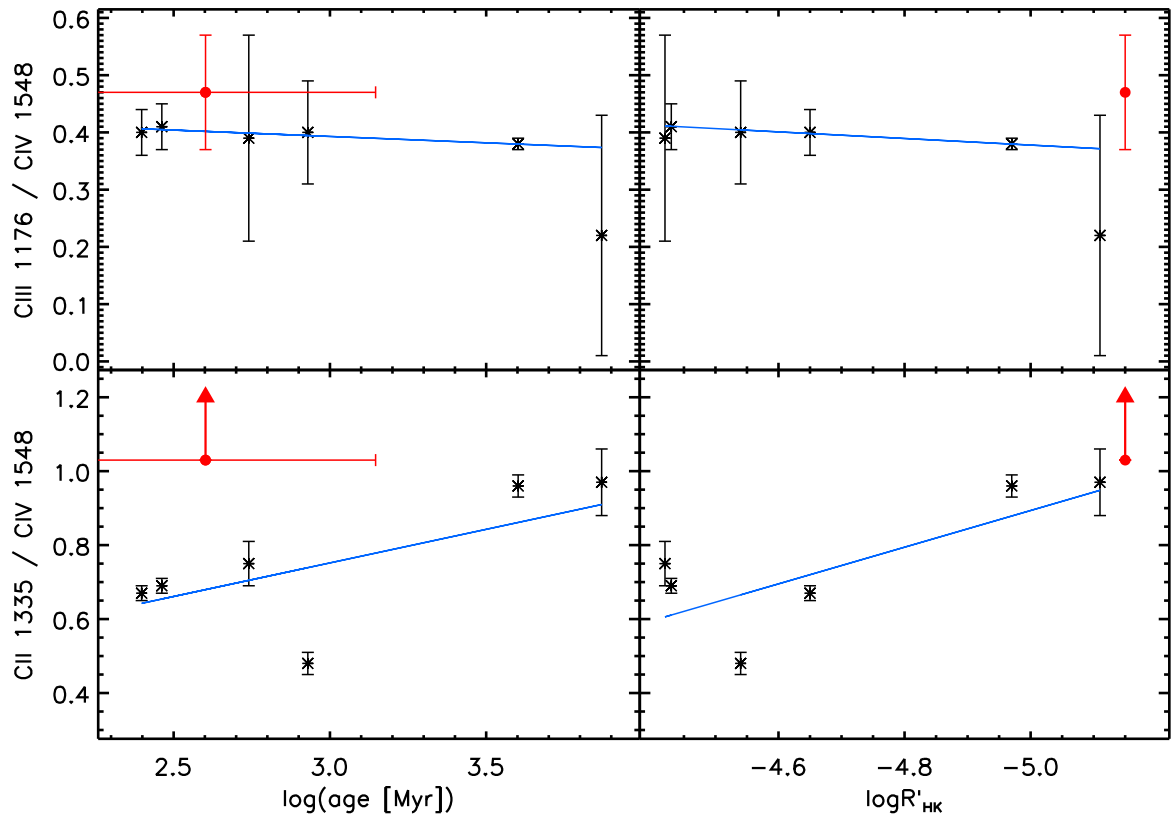}
\caption{Top panel: \ion{C}{3}/\ion{C}{4} flux ratio as a function of stellar age (left) and \logR\ value (right). WASP-18 is marked by a red circle, while the comparison stars are marked by black asterisks. The blue solid line shows the linear fit obtained by fitting the position of the comparisons stars. Bottom panel: same as the top panels, but for the \ion{C}{2}/\ion{C}{4} flux ratio. We remark that for WASP-18, because of the possible contamination by the ISM, the \ion{C}{2}/\ion{C}{4} flux ratio is a lower limit.}
\label{fig:ratio}
\end{figure*}

Figure~\ref{fig:ratio} shows the \ion{C}{3}/\ion{C}{4} (top) and \ion{C}{2}/\ion{C}{4} (bottom) flux ratios as a function of stellar age (left) and \logR\ value (right). We do not consider here the \ion{C}{4}/\ion{Si}{4} flux ratio because it does not vary significantly among the considered comparison stars and hence it is not constraining of the characteristics of WASP-18. We also looked at whether the upper limit on the \ion{C}{1} 1657\,\AA\ line flux provides meaningful constraints on the properties of WASP-18, but the lower limit on the \ion{C}{4}/\ion{C}{1} ratio of 0.65 allows one to conclude just that WASP-18 is younger than 6\,Gyr \citep[see Eq.~2 of][]{fossati2015b}. 

We find no significant trend in the \ion{C}{3}/\ion{C}{4} flux ratio as a function of age and \logR\ value, while the \ion{C}{2}/\ion{C}{4} flux ratio increases significantly with increasing age and \logR\ value. The two fits are
\begin{eqnarray}
\frac{\mathrm{CII}\,1335}{\mathrm{CIV}\,1548} &=& 0.207(\pm0.071) \nonumber \\
&+& 0.182(\pm0.026) \times \mathrm{log(Age\,[Myr])}
\label{eq:age}
\end{eqnarray}
and
\begin{eqnarray}
\frac{\mathrm{CII}\,1335}{\mathrm{CIV}\,1548} &=& -1.588(\pm0.278) \nonumber \\
&-& 0.496(\pm0.060) \times \mathrm{\log R^{\prime}_{\mathrm{HK}}}\,\,.
\label{eq:logR}
\end{eqnarray}
For the fits given in Eq.~(\ref{eq:age}) and (\ref{eq:logR}), we excluded WASP-18 and did not consider the uncertainties on age and \logR\ values.

Interestingly, the lower limit on the \ion{C}{2}/\ion{C}{4} flux ratio measured for WASP-18 is at the same level as that of the old, inactive comparison stars, in agreement with the low activity level inferred from the \ion{Ca}{2}\,H\&K lines and the X-ray fluxes.

To further confirm this result, we rescaled the fluxes measured for HD\,39587 and HD\,33262, the youngest stars in our sample, to the distance and radius of WASP-18. For the two comparison stars, we considered the distance listed in Table~\ref{tab:comp.stars} and a radius of 1.047 and 1.100 solar radii, respectively \citep{vanbelle2009,chandler2016}. We found that the FUV line fluxes measured for WASP-18 are, in general, about a factor of 10 smaller than expected if WASP-18 had the activity level of a young, active star like HD\,39587 and HD\,33262. 

Our results strongly support the postulated intrinsically low stellar activity, which \citet{pillitteri2014} suggested is due to the tidal influence of the massive planet (tidal star--planet interaction) affecting the outer stellar structure and activity of WASP-18. The presence of an intrinsically low stellar activity is further strengthened by comparing the measured upper limit on the X-ray fluxes with what expected on the basis of the stellar age. Following the relations of \citet{sanzforcada2011}, we obtain an upper limit on the EUV fluxes of \logleuv\,$<$\,27.9$\pm$0.2\,dex, where the uncertainty is caused by the scatter in the data-points, particularly at low activity levels, considered by \citet{sanzforcada2011}. We further converted this upper limit into a stellar age, obtaining a minimum age of about 9.6\,Gyr, which for a star like WASP-18 is longer than the main-sequence lifetime, thus in stark contrast with the star's young age. When using the age estimated from the X-ray luminosity vs age relationship of \citet{sanzforcada2011}, we would predict an X-ray luminosity of 1.93$\times$10$^{28}$\,erg\,s$^{-1}$, which is about 27 times larger than the upper limit derived from the {\it Chandra} observations.

\citet{Bonomoetal17} reconsidered all of the available observations of the WASP-18 system obtaining lower limits to the modified tidal quality factors of the star $Q^{\prime}_{\rm s}$ and of the planet $Q^{\prime}_{\rm p}$. We recall that the larger $Q^{\prime}$, the lower the dissipation of the tidal kinetic energy into the corresponding body \citep[see][for details]{Ogilvie14}. They found $Q^{\prime}_{\rm s} > 10^{8}$ and $Q^{\prime}_{\rm p} > 10^{7}$ in WASP-18 that implies that dynamical tides are likely not excited into the star, as suggested by \citet{OgilvieLin07}. The absence of a detectable tidal decay of the orbit of WASP-18b supports this conclusion \citep{Wilkinsetal17}. The consequence is that stellar rotation is probably not significantly affected by tides in this system and the remaining lifetime of the planet before it is engulfed by the star is at least of $\approx 100$\,Myr. Even if dynamical tides are likely not excited, the relatively large amplitude of the equilibrium tide in WASP-18 \citep[see][]{pillitteri2014} may imply a perturbation of the dynamo by the associated tidal flow. For example, it could increase the turbulence in the convection zone thus accelerating the decay of the magnetic fields in a stellar dynamo. A sufficiently large turbulent dissipation may quench a large-scale solar-like dynamo leaving only a small-scale turbulent dynamo that can sustain stellar activity only at the basal level. Nevertheless, this is just a qualitative conjecture that requires detailed physical modeling to be proven.
\subsection{WASP-18's high-energy flux}\label{sec:xuv}
The evolution of the atmosphere of a planet is mostly driven by atmospheric escape \citep[e.g.][]{jin2014}, which, in turn, is controlled by the high-energy stellar irradiation (XUV: X-ray [1--100\,\AA] $+$ EUV [100--912\,\AA]). This is therefore one of the most critical parameters for exoplanet studies \citep[e.g.][]{ribas2005,sanzforcada2011,linsky2014,france2016}. For all stars other than the Sun, direct measurements of the EUV flux are not possible because of absorption by neutral hydrogen in the ISM. To circumvent this problem, various indirect methods and scaling relations have been developed to estimate the XUV flux on the basis of accessible observables, such as X-ray fluxes \citep{sanzforcada2011,chadney2015}, Ly$\alpha$ fluxes \citep{linsky2014}, and stellar rotational velocities \citep{wood1994,colin2015,tu2015}.

\begin{figure}[h!]
\includegraphics[width=\hsize]{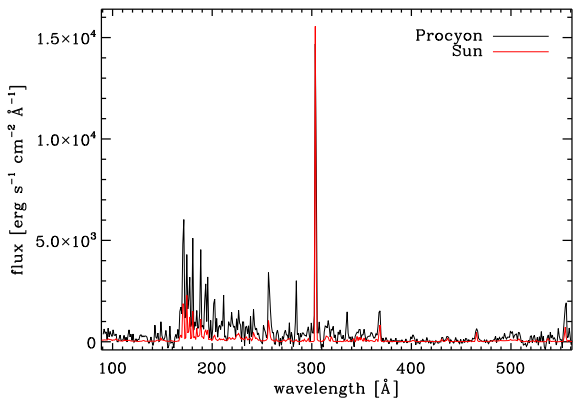}
\caption{Comparison between the EUV fluxes  at the stellar surface of Procyon (black) and the Sun (red).}
\label{fig:procyon}
\end{figure}

As detailed above, no adequate proxy for the high-energy spectrum of WASP-18 is available. X-ray and Ly$\alpha$ fluxes are not available, and the stellar age does not correlate with the fluxes measured in the X-rays (corona) and in the FUV emission lines (chromosphere). To estimate the stellar XUV fluxes, we can only follow the procedure we previously adopted for WASP-13 \citep{fossati2015b}. We rescaled the solar irradiance reference spectrum \citep{woods2009} to match the observed WASP-18 flux in the \ion{Si}{4} $\lambda$1394 feature, accounting for WASP-18's distance and radius. This \ion{Si}{4} feature is detected at a significance higher than that of the \ion{C}{4} $\lambda$1548 feature. By integrating the rescaled solar spectrum, we obtained an XUV (10--1180\,\AA) flux at the planetary distance of 10.2\,\mWm\  (or \ergscm), which is about twice that of the average Sun \citep{ribas2005}. This result is supported by a comparison of the observed EUV (100-600\,\AA) fluxes of Procyon, which has a spectral type F5V (hence close to that of WASP-18), and the Sun (Fig.~\ref{fig:procyon}) from which we find that the integrated EUV flux of Procyon is about three times larger than solar. We further inferred the X-ray luminosity from the scaled solar fluxes obtaining a value of 1.2$\times$10$^{26}$\,erg\,s$^{-1}$, which is about a factor of six smaller than the upper limit derived with {\it Chandra}, thus in agreement with the observations.
%
\subsection{Planet atmospheric escape}\label{sec:escape}
We use the model of \citet{koskinen2013a,koskinen2013b} to obtain the mass-loss rate from WASP-18b. We calculate the basic atmospheric structure to set the lower boundary conditions of our escape model by assuming an isothermal lower atmosphere with a dayside temperature of 3150\,K\footnote{The use of the temperature derived on the basis of the GAIA distance (Sect.~\ref{sec:sed}) has not effect on the results.} \citep{nymeyer2011,sheppard2017}. This is a rather extreme choice that leads to significant dissociation of H$_2$ above the 0.1\,bar level. In reality, somewhat more moderate temperature profiles fit the secondary eclipse data too, leading to less efficient dissociation and hence a less extended upper atmosphere \citep{iro13}. A more detailed exploration of the thermal structure is required to properly resolve the conditions in the middle atmosphere between $\sim$10\,mbar and 1\,$\mu$bar. A model of thermal structure applicable to WASP-18b in this region of the atmosphere, however, does not currently exist and the exploration is well beyond the scope of the present work. Fortunately, accurate modeling of the middle atmosphere is not required for our purposes. As we demonstrate below, the escape rate from WASP-18b is very low and this conclusion will not be changed by detailed radiative transfer models that instead are likely to produce even lower escape rates. 

In addition to the temperature, the atmospheric structure depends on the composition through the mean molecular weight. We assumed a solar He abundance \citep{lodders2003} and that the relative abundances of H$_2$ and H are in thermochemical equilibrium. The presence of trace species such as H$_2$O and CO in solar abundance does not significantly affect the mean molecular weight. Once the mean molecular weight as a function of pressure is known, the altitudes are obtained from the hypsometric equation
\begin{equation}
\int_{h_0}^{h}\frac{m(h')g(h'){\rm d}h'}{kT(h')}=-\ln\left[\frac{p(h)}{p(h_0)}\right]\,,
\end{equation}
where $m$ is the mean molecular weight, $k$ is the Boltzmann constant, $g$ is gravity, $T$ is temperature, $p$ is pressure, and the altitude coordinate $h=h(r,\theta,\phi)$ is measured perpendicular to surfaces of constant gravitational potential at a given location \citep[e.g.][]{koskinen2015a}.

The distortion of the planet by stellar gravity (i.e. the ``Roche lobe'' effect) and rotation are negligible (the L1 point is at 5.1\,$R_{\rm p}$) and thus we simply treat the planet and its gravitational potential as spherical, i.e. $h=r$. We set the lower boundary of the escape model at 1\,$\mu$bar where the altitude based on our structure model is 1452\,km and the mixing ratios of H$_2$, He, and H are 1.5$\times$10$^{-3}$, 7.35$\times$10$^{-2}$, and 0.925, respectively. We need to specify species densities, temperature, and the bulk velocity at the lower boundary. We do this by fixing the pressure and use it with the lower boundary temperature to calculate the total density and obtain species densities by fixing the lower boundary mixing ratios. We use mass flux conservation to set the lower boundary velocity during the simulation.         

\begin{figure}[h!]
\begin{center}
\includegraphics[width=\hsize,clip]{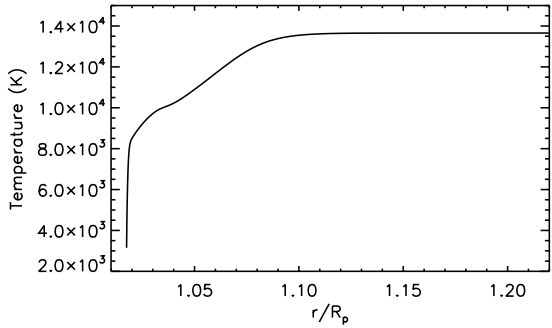}
\caption{Temperature profile of the upper atmosphere of WASP-18b based on our simulation.}
\label{fig:tp}
\end{center}
\end{figure}

Given a composition dominated by atomic hydrogen, the temperature begins to increase with altitude immediately above the lower boundary.  Figure~\ref{fig:tp} shows the temperature profile calculated by the model. We note here that our calculation also includes cooling by Ly$\alpha$ emissions introduced by \citet{murray2009} and later studied in detail by \citet{menager2013}. The temperature increases steeply in the lower atmosphere, reaching an isothermal value of 13,660\,K below the exobase, which is located at a relatively low radial distance of 1.22\,$R_{\rm p}$ with a pressure of 1.4$\times$10$^{-13}$\,bar. Even at such a high temperature, the atmosphere does not expand to high altitudes and, because of the large planetary mass, the Jeans escape parameter for H remains large $X_{\rm H}(r_{\rm exo})$\,=\,70.9. This shows that the upper atmosphere of WASP-18b is not in a hydrodynamic regime, which would require values of the Jeans parameter for H around 2--3 \citep{volkov2011}. The escape lies therefore in a Jeans escape regime, where the atmospheric particles move with a speed controlled by a Maxwellian distribution, while in the hydrodynamic regime the velocity distribution function evolves to a drifting Maxwellian that incorporates the bulk flow and the exobase extends to very high altitudes.

\begin{figure}[h!]
\includegraphics[width=\hsize,clip]{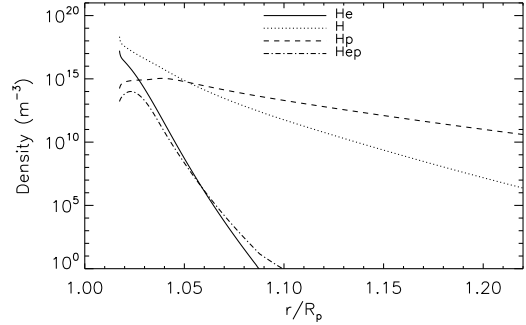}
\caption{Basic WASP-18b upper atmospheric composition based on our simulation. The solid and dotted lines are for neutral hydrogen and helium, respectively, while the dashed and dashed--dotted lines are for the respective singly ionized species.}
\label{fig:tsis}
\end{figure}

Figure~\ref{fig:tsis} shows the composition of the thermosphere-ionosphere on WASP-18b. Not surprisingly, the upper atmosphere is strongly ionized and H$^+$ overtakes as the dominant species above 1.05\,$R_{\rm p}$ ($p$\,=\,2.4\,nbar). Exceptionally for hot Jupiters, escape is negligible and the abundance of He and He$^+$ decreases with altitude more rapidly than that of H and H$^+$ due to diffusive separation. This is one of the few well known hot Jupiters to date where diffusive separation is relevant -- in most other cases, hydrodynamic escape leads to near uniform mixing of the species of interest. It means that the escape of heavy atoms and ions from the atmosphere of WASP-18b is even less likely than that of hydrogen.

As described above, the atmosphere of WASP-18b does not escape hydrodynamically due to the high mass of the planet, while Jeans escape holds. We treat this by imposing kinetic Jeans escape upper boundary conditions at the exobase (where the Knudsen number based on all collisions, evaluated and adjusted during the simulation, is $Kn$\,=\,1). The bulk outflow velocity at the upper boundary is based on the sum of the Jeans outflow fluxes for each species. In addition, we impose upper boundary temperature and species density gradients consistent with Jeans escape. In effect, we adapt the Type-I upper boundary conditions from \citet[][equations (14)--(19)]{zhu2014} that were generalized for multiple species by \citet{koskinen2015b}.

As the upper atmosphere of WASP-18b is strongly ionized, we follow \citet{koskinen2013a} in writing the thermal escape (Jeans) parameter for ions and electrons (species $s$) as
\begin{equation}
X_{\rm s}(r_{\rm exo}) = X_{\rm sg}(r_{\rm exo}) + X_{\rm se}(r_{\rm exo}) = \frac{\Delta\phi_{\rm exo}m_{\rm s}}{kT_{\rm exo}} - \frac{q_{\rm s}\phi_{\rm e}(r_{\rm exo})}{kT_{\rm exo}}\,,
\end{equation}
where $\phi_{\rm e}$ is the ambipolar electric potential, $q_{\rm s}$ is the charge of the ion/electron, $\Delta\phi_{\rm exo}$ is the gravitational potential difference from the exobase to the Roche lobe, $T_{\rm exo}$ is the temperature at the exobase, $r_{\rm exo}$ is the exobase radius, and $X_{\rm sg}$ and $X_{\rm se}$ are the gravitational and electrostatic escape parameters, respectively. The value for $\phi_{\rm e}(r_{\rm exo})$ is obtained iteratively by imposing the condition of zero electric current through the upper boundary. As electrons are much lighter than ions, they escape more readily and drag ions along. The ambipolar electric field therefore has the effect of accelerating ion escape (decreasing $X_{\rm s}$ for positive $q_{\rm s}$) while slowing down electron escape (increasing $X_{\rm s}$ for electrons i.e. $q_{\rm s}$\,$<$\,0) until the zero electric current condition is satisfied. Note that our upper boundary conditions, as indicated, also account for stellar gravity (i.e. ``Roche lobe effect'') although this is negligible for WASP-18b. 

It is interesting to explore the effect of ambipolar electric fields on $X_{\rm s}$, even though in this case the effect does not lead to significantly enhanced mass loss. The direct effect of electric fields is of course negligible for neutrals (although collisions with ions can impart some acceleration on them) and the abundance of He$^+$ near the exobase is negligible. Thus, we focus on H and H$^+$. As stated above, $X_{\rm H}(r_{\rm exo})$\,=\,70.9. Our calculation shows that $X_{\rm H^+e}(r_{\rm exo})$\,=\,37.4 and thus $X_{\rm H^+}(r_{\rm exo})$\,=\,33.5 for H$^+$. A Jeans parameter $X_{\rm s}$ of 33.5, however, is still high and the bulk outflow velocity at the exobase is only 4$\times$10$^{-10}$\,\ms, giving a negligible mass loss rate of 2.8$\times$10$^{-9}$\,\kgs\  (8.8$\times$10$^7$\,kg\,Gyr$^{-1}$ or 4.6$\times$10$^{-20}$\,$M_{\rm J}$\,Gyr$^{-1}$). Thus mass loss from WASP-18b is not energy limited. Energy-limited mass loss would be inversely proportional to planet mass, whereas Jeans escape is instead proportional to $(1 + X_{\rm s})e^{-X_{\rm s}}$ and thus the rate decreases very rapidly with increasing $X_{\rm s}$, or planet mass.

To strengthen this result, we calculate the exobase temperature required for the atmosphere to be in the hydrodynamic escape regime and thus dramatically increase the escape rates. The approximate temperature can be directly inferred from the Jeans escape parameter\footnote{We remind that the Jeans escape parameter decreases linearly with increasing temperature.} considering that hydrodynamic escape requires the Jeans escape parameter to be below two. Figure~\ref{fig:tp} shows that the upper atmosphere has a temperature of about 1.3$\times$10$^4$\,K. To reach a Jeans escape parameter of about two, the temperature should therefore be of the order of 2.2$\times$10$^5$\,K, which is impossible to reach through heating from the stellar EUV flux. This further indicates that even much larger EUV fluxes would not significantly affect the atmospheric escape of WASP-18b.

The presence of an enhanced atmospheric metallicity for WASP-18b \citep{sheppard2017} would not significantly affect our results. A higher metallicity would produce an even more compact atmosphere, which is less likely to escape. Given that our pure H/He atmosphere presents a very low escape rate, the high metallicity derived by \citet{sheppard2017} is most likely to be primordial and not caused by selective erosion (i.e. lighter elements are more likely to escape compared to heavier ones), even considering that the star may have been more active in the past.

\section{Conclusions}\label{sec:conclusions}
We collected and analyzed FUV COS/{\it HST} spectra of WASP-18, which hosts a massive hot Jupiter in a very close orbit. This star is particularly interesting because, despite being younger than 1\,Gyr, it presents a very low activity level. In fact, no chromospheric/coronal emission is observed at the core of the \ion{Ca}{2}\,H\&K resonance lines or in X-rays.

Extrinsic absorption from the ISM is a possible explanation for the observed anomalously low activity level. For this reason, we inferred the ISM column densities for \ion{H}{1}, \ion{C}{1}, \ion{C}{2}, \ion{C}{3}, \ion{C}{4}, \ion{Si}{4}, \ion{Mg}{2}, \ion{Ca}{2}, and \ion{Na}{1} in the direction of WASP-18, concluding that ISM absorption is not the cause of the anomaly. This agrees with inferences from the WASP-18 X-ray observations by \citet{pillitteri2014}.

We further measured the four stellar emission features identified in the COS spectrum. We clearly detected the emission of the \ion{C}{3} multiplet at 1176\,\AA, the \ion{C}{2} doublet at 1335\,\AA, the \ion{Si}{4} line at 1394\,\AA, and the \ion{C}{4} doublet at 1548\,\AA.

We then compared the \ion{C}{3}/\ion{C}{4} and \ion{C}{2}/\ion{C}{4} flux ratios obtained for WASP-18 (the latter as a lower limit) with those derived in the same manner from nearby stars with spectral type similar to that of WASP-18 and with known age. We found that the \ion{C}{3}/\ion{C}{4} flux ratio does not vary significantly with age, while the \ion{C}{2}/\ion{C}{4} flux ratio increases with age or equivalently with decreasing activity level. Interestingly, despite its young age, WASP-18 presents a \ion{C}{2}/\ion{C}{4} flux ratio comparable to that of old, inactive stars. This confirms that the discrepancy between the star's age and activity level is indeed caused by an intrinsically low stellar activity. \citet{pillitteri2014} suggested this could be caused by tidal interaction between the massive planet and the star.

Having no better option, we inferred the stellar XUV flux by re-scaling the solar irradiance reference spectrum on the basis of the measured FUV fluxes. We obtained a high-energy flux about twice that of the average Sun, in agreement with the comparison of the EUV spectra of the Sun and Procyon, the latter being not too dissimilar from WASP-18. We used these XUV fluxes to model the planetary upper atmosphere, thus estimating the mass-loss rate. Because of the large planetary mass, we found that the mass-loss rate is negligible in terms of its effect on planetary evolution. Mass loss is mostly driven by Jeans escape, hence it is not energy limited. The possible metal rich nature of the planetary atmosphere \citep{sheppard2017} is therefore likely to be primordial.

We have demonstrated the power of FUV stellar spectroscopy in identifying the cause of anomalous activity levels, a peculiarity which is found among 25\% of planet-hosting stars \citep{staab2017}. Observations like these are possible exclusively thanks to the high throughput of the COS optical system and the low background of its detector. By following similar designs and characteristics, future UV instrumentation for both large \citep[e.g. HABEX, LUVOIR;][]{france2017} and small \citep[e.g. CUTE;][]{fleming2017} space telescopes will allow us to look deeper into star--planet interactions and eventually to pin down the mechanisms operating through the observed behavior of a wide variety of systems.
\section*{Acknowledgments}
This work is based on observations made with the NASA/ESA {\it Hubble
Space Telescope}, obtained from MAST at the Space Telescope Science Institute,
which is operated by the Association of Universities for Research in
Astronomy, Inc., under NASA contract NAS 5-26555. These observations are
associated with program No. 13859, to which support was provided by NASA through
a grant from the Space Telescope Science Institute. Support for StarCAT was provided by grant HST-AR-10638.01-A from STScI and grant NAG5-13058 from NASA. Based on observations made with ESO Telescopes at the La Silla Paranal Observatory under programme ID 092.D-0587. This work has made use of public databases hosted by SIMBAD and VizieR, both maintained by CDS, Strasbourg, France. We thank the anonymous referee for the very useful and constructive comments. C.H. is supported by STFC under grant ST/P000584/1. I.P. acknowledges support from INAF and ASI through the ARIEL consortium.
{\it Facilities:} \facility{HST (COS, STIS)}, \facility{VLT (UVES)}.


\begin{thebibliography}{}
\bibitem[Am{\^o}res \& L{\'e}pine(2005)]{amores2005}	
	Am{\^o}res, E.~B., \& L{\'e}pine, J.~R.~D. 2005, \aj, 130, 679
\bibitem[Ayres(2010)]{ayres2010}
	Ayres, T.~R.\ 2010, \apjs, 187, 149
\bibitem[Baliunas \& Jastrow(1990)]{baliunas1990}
	Baliunas, S. \& Jastrow, R.\ 1990, \nat, 348, 520 
\bibitem[Barnes(2007)]{barnes2007}
	Barnes, S.~A.\ 2007, \apj, 669, 1167 
\bibitem[Beaug{\'e} \& Nesvorn{\'y}(2013)]{beauge2013}
	Beaug{\'e}, C. \& Nesvorn{\'y}, D.\ 2013, \apj, 763, 12 
\bibitem[Bessell et al.(1998)]{bessel1998} 
	Bessell, M.~S., Castelli, F., \& Plez, B.\ 1998, \aap, 333, 231 
\bibitem[Bonomo et al.(2017)]{Bonomoetal17}
	Bonomo, A.~S., Desidera, S., Benatti, S., et al.\ 2017, \aap, 602, A107
\bibitem[Bourrier et al.(2013)]{bourrier2013}
	Bourrier, V., Lecavelier des Etangs, A., Dupuy, H., et al.\ 2013, \aap, 		551, A63 
\bibitem[Bressan et al.(2012)]{bressan2012} 
	Bressan, A., Marigo, P., Girardi, L., et al.\ 2012, \mnras, 427, 127 
\bibitem[Carroll-Nellenback et al.(2016)]{carroll2016}
	Carroll-Nellenback, J., Frank, A., Liu, B., et al.\ 2017, \mnras, 466, 			2458
\bibitem[Casagrande et al.(2011)]{casagrande2011}
	Casagrande, L., Sch{\"o}nrich, R., Asplund, M., et al. 2011, \aap, 530, 		A138 
\bibitem[Chabrier(2001)]{chabrier2001} 
	Chabrier, G.\ 2001, \apj, 554, 1274 
\bibitem[Chadney et al.(2015)]{chadney2015}
	Chadney, J.~M., Galand, M., Unruh, Y.~C., Koskinen, T.~T. \& 				Sanz-Forcada, J. 2015, \icarus, 250, 357 
\bibitem[Chandler et al.(2016)]{chandler2016} 
	Chandler, C.~O., McDonald, I., \& Kane, S.~R.\ 2016, \aj, 151, 59 
\bibitem[Cohen et al.(2011)]{cohen2011} 
	Cohen, O., Kashyap, V.~L., Drake, J.~J., et al.\ 2011, \apj, 733, 67 
\bibitem[Cutri et al.(2003)]{cutri2003} 
	Cutri, R.~M., Skrutskie, M.~F., van Dyk, S., et al.\ 2003, VizieR Online 	Data Catalog, 2246
\bibitem[Cutri \& et al.(2012)]{cutri2012} 
	Cutri, R.~M., \& et al.\ 2012, VizieR Online Data Catalog, 2311  
\bibitem[Danforth et al.(2010)]{danforth2010}
	Danforth, C.~W., Keeney, B.~A., Stocke, J.~T., Shull, J.~M., \& Yao, Y.\ 	2010, \apj, 720, 976  
\bibitem[Davis \& Wheatley(2009)]{davis2009} 
	Davis, T.~A., \& Wheatley, P.~J.\ 2009, \mnras, 396, 1012 
\bibitem[Diplas \& Savage(1994)]{diplas1994}
	Diplas, A. \& Savage, B.~D. 1994, \apj, 427, 274 
\bibitem[Donahue(1998)]{donahue1998}
	Donahue, R.~A.\ 1998, Cool Stars, Stellar Systems, and the Sun, 154, 			1235 
\bibitem[Duncan et al.(1991)]{duncan1991}
	Duncan, D.~K., Vaughan, A.~H., Wilson, O.~C., et al.\ 1991, \apjs, 76, 			383
\bibitem[Ehrenreich et al.(2015)]{ehrenreich2015} 
        Ehrenreich, D., Bourrier, V., Wheatley, P.~J., et al.\ 2015, \nat, 522, 		459
\bibitem[Figueira et al.(2014)]{figueira2014}
	Figueira, P., Oshagh, M., Adibekyan, V.~Z. \& Santos, N.~C. 2014, \aap, 		572, A51 
\bibitem[Fleming et al.(2017)]{fleming2017}
	Fleming, B., France, K., Nell, N., et al.\ 2017, \procspie, 10397, 			1039711
\bibitem[Fossati et al.(2010)]{fossati2010}
	Fossati, L., Haswell, C.~A., Froning, C.~S., et al.\ 2010, \apjl, 714, 			L222 
\bibitem[Fossati et al.(2013)]{fossati2013}
	Fossati, L., Ayres, T.~R., Haswell, C.~A., et al.\ 2013, \apjl, 766, L20
\bibitem[Fossati et al.(2014)]{fossati2014}
	Fossati, L., Ayres, T.~R., Haswell, C.~A., et al.\ 2014, \apss, 354, 21 
\bibitem[Fossati et al.(2015a)]{fossati2015a}
	Fossati, L., Haswell, C.~A., Linsky, J.~L. \& Kislyakova, K.~G. 2015a, 			Astrophysics and Space Science Library, Vol. 411, 59 
\bibitem[Fossati et al.(2015b)]{fossati2015b}
	Fossati, L., France, K., Koskinen, T., et al.\ 2015b, \apj, 815, 118 
\bibitem[Fossati et al.(2015c)]{fossati2015c}
	Fossati, L., Ingrassia, S. \& Lanza, A.~F.\ 2015c, \apjl, 812, L35
\bibitem[Fossati et al.(2017)]{fossati2017}
	Fossati, L., Marcelja, S.~E., Staab, D., et al.\ 2017, \aap, 601, A104
\bibitem[France et al.(2010)]{france2010} 
	France, K., Stocke, J.~T., Yang, H., et al.\ 2010, \apj, 712, 1277 
\bibitem[France et al.(2013)]{france2013}
	France, K., Froning, C.~S., Linsky, J.~L., et al.\ 2013, \apj, 763, 149 
\bibitem[France et al.(2016)]{france2016} 
	France, K., Parke Loyd, R.~O., Youngblood, A., et al.\ 2016, \apj, 820, 		89 
\bibitem[France et al.(2017)]{france2017} 
	France, K., Fleming, B., West, G., et al.\ 2017, \procspie, 10397, 			1039713
\bibitem[Frisch \& Slavin(2003)]{frisch2003}
	Frisch, P.~C. \& Slavin, J.~D. 2003, \apj, 594, 844 
\bibitem[Frisch et al.(2011)]{frisch2011} 
	Frisch, P.~C., Redfield, S., \& Slavin, J.~D.\ 2011, \araa, 49, 237 
\bibitem[Gaia Collaboration et al.(2016)]{tgas} 
	Gaia Collaboration, Prusti, T., de Bruijne, J.~H.~J., et al.\ 2016, 			\aap, 595, A1 
\bibitem[Gaudi et al.(2017)]{gaudi2017} 
	Gaudi, B.~S., Stassun, K.~G., Collins, K.~A., et al.\ 2017, \nat, 546, 			514 
\bibitem[Green et al.(2012)]{green2012}
	Green, J.~C., Froning, C.~S., Osterman, S., et al.\ 2012, \apj, 744, 60 
\bibitem[Grossmann et al.(1995)]{grossmann1995} 
	Grossmann, V., Baessgen, G., Evans, D.~W., et al.\ 1995, \aap, 304, 110 
\bibitem[Gustafsson et al.(2008)]{marcs}
	Gustafsson, B., Edvardsson, B., Eriksson, K., et al. 2008, \aap, 486, 			951 
\bibitem[G{\"u}ver \& \"Ozel(2009)]{guver2009}
	G{\"u}ver, T. \& \"Ozel, F. 2009, \mnras, 400, 2050 
\bibitem[Haswell(2010)]{haswell2010} 
	Haswell, C.~A.\ 2010, Transiting Exoplanets by Carole 					A.~Haswell.~Cambridge University Press, 2010.~ISBN: 9780521139380,  
\bibitem[Haswell et al.(2012)]{haswell2012}
	Haswell, C.~A., Fossati, L., Ayres, T., et al.\ 2012, \apj, 760, 79
\bibitem[Haswell(2017)]{haswell2017} 
	Haswell, C.~A.\ 2017, Handbook of Exoplanets, arXiv:1708.00693 
\bibitem[Hebb et al.(2009)]{hebb2009}
	Hebb, L., Collier-Cameron, A., Loeillet, B., et al.\ 2009, \apj, 693, 			1920 
\bibitem[Hellier et al.(2009)]{hellier2009}
	Hellier, C., Anderson, D.~R., Collier Cameron, A., et al.\ 2009, \nat, 			460, 1098 
\bibitem[H{\o}g et al.(2000)]{hog2000} 
	H{\o}g, E., Fabricius, C., Makarov, V.~V., et al.\ 2000, \aap, 355, L27
\bibitem[Iro \& Maxted(2013)]{iro13}
        Iro, N. \& Maxted, P. F. L. \ 2013, \icarus, 226, 1719
\bibitem[Jin et al.(2014)]{jin2014} 
	Jin, S., Mordasini, C., Parmentier, V., et al.\ 2014, \apj, 795, 65 
\bibitem[Jin \& Mordasini(2017)]{jin2017} 
	Jin, S., \& Mordasini, C.\ 2017, ApJ, in press (arXiv:1706.00251) 
\bibitem[Johnstone et al.(2015)]{colin2015}
	Johnstone, C.~P., G{\"u}del, M., Brott, I. \& L\"uftinger, T. 2015, 			\aap, 577, A28 
\bibitem[Keil et al.(1998)]{keil1998}
	Keil, S.~L., Henry, T.~W. \& Fleck, B.\ 1998, Synoptic Solar Physics, 			Vol. 140, 301 
\bibitem[Kharchenko et al.(2009)]{kharchenko2009} 
	Kharchenko, N.~V., Piskunov, A.~E., R{\"o}ser, S., et al.\ 2009, \aap, 			504, 681 
\bibitem[Koskinen et al.(2013a)]{koskinen2013a}
	Koskinen, T.~T., Harris, M.~J., Yelle, R.~V. \& Lavvas, P.\ 2013a, 			\icarus, 226, 1678 
\bibitem[Koskinen et al.(2013b)]{koskinen2013b}
	Koskinen, T.~T., Yelle, R.~V., Harris, M.~J. \& Lavvas, P.\ 2013b, 			\icarus, 226, 1695 
\bibitem[Koskinen et al.(2015a)]{koskinen2015a} 
	Koskinen, T.~T., Sandel, B.~R., Yelle, R.~V., et al.\ 2015a, \icarus, 			260, 174 
\bibitem[Koskinen et al.(2015b)]{koskinen2015b} 
	Koskinen, T.~T., Erwin, J.~T., \& Yelle, R.~V.\ 2015, \grl, 42, 7200
\bibitem[Lammer et al.(2003)]{lammer2003}
	Lammer, H., Selsis, F., Ribas, I., et al.\ 2003, \apjl, 598, L121 
\bibitem[Lanza(2014)]{lanza2014} 
	Lanza, A.~F.\ 2014, \aap, 572, L6 
\bibitem[Lecavelier des Etangs et al.(2004)]{lecav2004} 
	Lecavelier des Etangs, A., Vidal-Madjar, A., McConnell, J.~C., \& 			H{\'e}brard, G.\ 2004, \aap, 418, L1 
\bibitem[Linsky et al.(2010)]{linsky2010}
	Linsky, J.~L., Yang, H., France, K., et al.\ 2010, \apj, 717, 1291 
\bibitem[Linsky et al.(2013)]{linsky2013}
	Linsky, J.~L., France, K. \& Ayres, T.\ 2013, \apj, 766, 69 
\bibitem[Linsky et al.(2014)]{linsky2014}
	Linsky, J.~L., Fontenla, J. \& France, K. 2014, \apj, 780, 61 
\bibitem[Livingston et al.(2007)]{livingston2007}
	Livingston, W., Wallace, L., White, O.~R., \& Giampapa, M.~S.\ 2007, 			\apj, 657, 1137
\bibitem[Lodders(2003)]{lodders2003} 
	Lodders, K.\ 2003, \apj, 591, 1220 
\bibitem[Mamajek \& Hillenbrand(2008)]{mamajek2008}
	Mamajek, E.~E. \& Hillenbrand, L.~A.\ 2008, \apj, 687, 1264 
\bibitem[Matsakos et al.(2015)]{matsakos2015} 
	Matsakos, T., Uribe, A., {\ K\"o}nigl, A.\ 2015, \aap, 578, A6
\bibitem[Matsakos \& K{\"o}nigl(2016)]{matsakos2016} 
	Matsakos, T., \& K{\"o}nigl, A.\ 2016, \apjl, 820, L8 
\bibitem[Maxted et al.(2013)]{maxted2013} 
	Maxted, P.~F.~L., Anderson, D.~R., Doyle, A.~P., et al.\ 2013, \mnras, 			428, 2645 
\bibitem[Mazeh et al.(2016)]{mazeh2016} 
	Mazeh, T., Holczer, T., \& Faigler, S.\ 2016, \aap, 589, A75 
\bibitem[Melo et al.(2006)]{melo2006}
	Melo, C., Santos, N.~C., Pont, F., et al. 2006, \aap, 460, 251 
\bibitem[Menager et al.(2013)]{menager2013} 
	Menager, H., Barth{\'e}lemy, M., Koskinen, T., et al.\ 2013, \icarus, 			226, 1709 
\bibitem[Miller et al.(2012)]{miller2012}
	Miller, B.~P., Gallo, E., Wright, J.~T., \& Dupree, A.~K.\ 2012, \apj, 			754, 137 
\bibitem[Mordasini et al.(2012)]{mordasini2012}
	Mordasini, C., Alibert, Y., Georgy, C., et al.\ 2012, \aap, 547, A112 
\bibitem[Murray-Clay et al.(2009)]{murray2009} 
	Murray-Clay, R.~A., Chiang, E.~I., \& Murray, N.\ 2009, \apj, 693, 23 
\bibitem[Nichols et al.(2015)]{nichols2015} 
	Nichols, J.~D., Wynn, G.~A., Goad, M., et al.\ 2015, \apj, 803, 9 
\bibitem[Nymeyer et al.(2011)]{nymeyer2011} 
	Nymeyer, S., Harrington, J., Hardy, R.~A., et al.\ 2011, \apj, 742, 35 
\bibitem[Ogilvie(2014)]{Ogilvie14} 
	Ogilvie, G.~I.\ 2014, \araa, 52, 171 
\bibitem[Ogilvie \& Lin(2007)]{OgilvieLin07}
	Ogilvie, G.~I., \& Lin, D.~N.~C.\ 2007, \apj, 661, 1180
\bibitem[Owen \& Wu(2013)]{owen2013}
	Owen, J.~E., \& Wu, Y.\ 2013, \apj, 775, 105 
\bibitem[Owen \& Wu(2017)]{owen2017} 
	Owen, J.~E., \& Wu, Y.\ 2017, ApJ, in press (arXiv:1705.10810) 
\bibitem[Pace(2013)]{pace2013}
	Pace, G. 2013, \aap, 551, L8 
\bibitem[Pagano et al.(2004)]{pagano2004} 
	Pagano, I., Linsky, J.~L., Valenti, J., \& Duncan, D.~K.\ 2004, \aap, 			415, 331 
\bibitem[Perryman et al.(1997)]{perryman1997} 
	Perryman, M.~A.~C., Lindegren, L., Kovalevsky, J., et al.\ 1997, \aap, 			323, L49 
\bibitem[Pillitteri et al.(2014)]{pillitteri2014}
	Pillitteri, I., Wolk, S.~J., Sciortino, S. \& Antoci, V.\ 2014, \aap, 			567, A128 
\bibitem[Piters et al.(1997)]{piters1997} 
	Piters, A.~J.~M., Schrijver, C.~J., Schmitt, J.~H.~M.~M., et al.\ 1997, 		\aap, 325, 1115 
\bibitem[Rachford et al.(2002)]{rachford2002}
	Rachford, B.~L., Snow, T.~P., Tumlinson, J., et al. 2002, \apj, 577, 221
\bibitem[Radick et al.(1998)]{radick1998}
	Radick, R.~R., Lockwood, G.~W., Skiff, B.~A. \& Baliunas, S.~L.\ 1998, 			\apjs, 118, 239 
\bibitem[Redfield \& Linsky(2002)]{redfield2002}
	Redfield, S. \& Linsky, J.~L. 2002, \apjs, 139, 439 
\bibitem[Redfield \& Linsky(2004)]{redfield2004} 
	Redfield, S., \& Linsky, J.~L.\ 2004, \apj, 602, 776 
\bibitem[Ribas et al.(2005)]{ribas2005}
	Ribas, I., Guinan, E.~F., G{\"u}del, M. \& Audard, M. 2005, \apj, 622, 			680 
\bibitem[Sanz-Forcada et al.(2011)]{sanzforcada2011}
	Sanz-Forcada, J., Micela, G., Ribas, I., et al. 2011, \aap, 532, A6 
\bibitem[Savage \& Mathis(1979)]{savage1979}
	Savage, B.~D. \& Mathis, J.~S. 1979, \araa, 17, 73 
\bibitem[Sheppard et al.(2017)]{sheppard2017} 
	Sheppard, K.~B., Mandell, A.~M., Tamburo, P., et al.\ 2017, \apjl, 850, 		L32 
\bibitem[da Silva et al.(2006)]{dasilva2006} 
	da Silva, L., Girardi, L., Pasquini, L., et al.\ 2006, \aap, 458, 609 
\bibitem[Snow \& McCall(2006)]{snow2006}
	Snow, T.~P. \& McCall, B.~J. 2006, \araa, 44, 367 
\bibitem[Southworth(2012)]{southworth2012}
	Southworth, J.\ 2012, \mnras, 426, 1291 
\bibitem[Staab et al.(2017)]{staab2017}
	Staab, D., Haswell, C.~A., Smith, G.~D., et al.\ 2017, \mnras, 466, 738
\bibitem[Stelzer et al.(2013)]{stelzer2013}
	Stelzer, B., Marino, A., Micela, G., L{\'o}pez-Santiago, J. \& Liefke, 			C.\ 2013, \mnras, 431, 2063 
\bibitem[Szab{\'o} \& Kiss(2011)]{szabo2011} 
	Szab{\'o}, G.~M., \& Kiss, L.~L.\ 2011, \apjl, 727, L44 
\bibitem[Triaud et al.(2010)]{triaud2010}
	Triaud, A.~H.~M.~J., Collier Cameron, A., Queloz, D., et al.\ 2010, 			\aap, 524, A25 
\bibitem[Tu et al.(2015)]{tu2015}
	Tu, L., Johnstone, C.~P., G{\"u}del, M. \& Lammer, H. 2015, \aap, 577, 			L3 
\bibitem[van Belle \& von Braun(2009)]{vanbelle2009} 
	van Belle, G.~T., \& von Braun, K.\ 2009, \apj, 694, 1085 
\bibitem[van der Bliek et al.(1996)]{vander} 
	van der Bliek, N.~S., Manfroid, J., \& Bouchet, P.\ 1996, \aaps, 119, 			547 
\bibitem[van Leeuwen(2007)]{vanLeeuwen2007}
	van Leeuwen, F. 2007, \aap, 474, 653 
\bibitem[Vidal-Madjar et al.(2003)]{vidal2003}
	Vidal-Madjar, A., Lecavelier des Etangs, A., D{\'e}sert, J.-M., et al.\ 		2003, \nat, 422, 143 
\bibitem[Volkov et al.(2011)]{volkov2011} 
	Volkov, A.~N., Johnson, R.~E., Tucker, O.~J., \& Erwin, J.~T.\ 2011, 			\apjl, 729, L24 
\bibitem[Welsh et al.(1994)]{welsh1994} 
	Welsh, B.~Y., Craig, N., Vedder, P.~W., \& Vallerga, J.~V.\ 1994, \apj, 		437, 638 
\bibitem[Welsh et al.(1997)]{welsh1997} 
	Welsh, B.~Y., Sasseen, T., Craig, N., Jelinsky, S., \& Albert, C.~E.\ 			1997, \apjs, 112, 507 
\bibitem[Welsh et al.(2010)]{welsh2010}
	Welsh, B.~Y., Wheatley, J., Siegmund, O.~H.~W. \& Lallement, R. 2010, 			\apjl, 712, L199 
\bibitem[Wilkins et al.(2017)]{Wilkinsetal17}
	Wilkins, A.~N., Delrez, L., Barker, A.~J., et al.\ 2017, \apjl, 836, L24
\bibitem[Wood et al.(1994)]{wood1994}
	Wood, B.~E., Brown, A., Linsky, J.~L., et al. 1994, \apjs, 93, 287 
\bibitem[Woods et al.(2009)]{woods2009}
	Woods, T.~N., Chamberlin, P.~C., Harder, J.~W., et al. 2009, \grl, 36, 			L01101 
\bibitem[Wright et al.(2004)]{wright2004}
	Wright, J.~T., Marcy, G.~W., Butler, R.~P. \& Vogt, S.~S.\ 2004, \apjs, 		152, 261 
\bibitem[Wright et al.(2010)]{wright2010} 
	Wright, E.~L., Eisenhardt, P.~R.~M., Mainzer, A.~K., et al.\ 2010, \aj, 		140, 1868 
\bibitem[Yelle(2004)]{yelle2004}
	Yelle, R.~V.\ 2004, \icarus, 170, 167 
\bibitem[Zhu et al.(2014)]{zhu2014} 
	Zhu, X., Strobel, D.~F., \& Erwin, J.~T.\ 2014, \icarus, 228, 301 
\end{thebibliography}
\end{document}